\documentstyle[11pt]{article}
\def\be{\begin{equation}}
\def\bea{\begin{eqnarray}}
\def\eea{\end{eqnarray}}
\def\l{\label}
\def\r{\ref}

\def\ee{\end{equation}}

\def\c{\cite}
\begin{document}
\begin{center}
\LARGE {\bf More on Phase Structure of   Nonlocal 2D Generalized
Yang-Mills Theories (nlgYM$_2$'s)}
\\
\end{center}
 {\bf $^{\dag, \ddag}$Kh. Saaidi \footnote{ E-mail: ksaaidi@hotmail.com},
 $^{\dag, \ddag}$ M.R. Setare  \footnote{E-mail: rezakord@yahoo.com}
 }\\
 {\it${^{\dag}}$ Department of Science, University of Kurdistan  ,
 Pasdaran Ave., Sanandaj, Iran} \\
 {\it $ ^{\ddag}$ Institute for Studies in Theoretical Physics and Mathemaics,
 P.O.Box, 19395-5531, Tehran, Iran}\\
\vskip 3cm
\begin{center}
{\bf{Abstract}}\\
\end{center}

We  study the phase structure of nonlocal two dimensional
generalized Yang - Mills theories (nlgYM$_2$) and it is shown that
all order of $\phi^{2k}$ model of these theories has phase
transition only on compact manifold with $g = 0$( on sphere), and
the order of phase transition is 3. Also it is shown that the
$\phi^2 + \frac{2\alpha}{3}\phi^3$ model of nlgYM$_2$ has third
order phase transition on any compact manifold with $1 < g < 1+
\frac{\hat{A}}{|\eta_c|}$, and has no phase transition on sphere.

\newpage
{\section{Introduction}}

The two dimensional Yang-Mills theory ($YM_2$) is a theoretical
laboratory for understanding the main theory of particle physics,
$QCD_4$. In recent years there have been much effort  to analyze
the different aspects of this theory [1-8]. It is well known that
$YM_2$ is defined by the Lagrangian ${\rm tr}(\frac{1}{4}F^2)$ on
a Riemann surface where $F$ is the field strength  tensor.
 This theory have certain properties, such as
invariance under area preserving diffeomorphism and lack of any
propagating degrees of freedom \c{b1}. In a $YM_2$ one starts from
a B-F theory in which a Lagrangian of the form $i{\rm tr}(BF) +
{\rm tr}(B^2)$ is used. There are, however, the many way to
generalized these theories without losing properties. One way is
so- called generalized Yang - Mills theory (gYM$_2$'s). These
theories are defined by replacing the term ${\rm tr}(B^2)$ in the
YM$_2$ Lagrangian by an arbitrary class function of $B$\c{e1}. It
is worthy of mention  that for gYM$_2$, one can not eliminate the
auxiliary field that obtain a Lagrangian for the gauge field.
Remarkable that the large gauge group limit of YM$_2$ and gYM$_2$
theories is also interesting. Several aspects of this theory have
been studied in [13 - 19]. There is another way to generalize
YM$_2$ and gYM$_2$ and that is to use a nonlocal action for the
auxiliary field, leading to the so - called nonlocal YM$_2$
(nlYM$_2$) and nonlocal gYM$_2$(nlgYM$_2$) theories, respectively
\c{kh1}. It is remarkable that,  the action of nlYM$_2$ and
nlgYM$_2$ is no extensive \c{kh1}. Several aspects of nlgYM$_2$,
such as, wave function, partition function, generating functional,
and also large $N$ limit of it, have been studied on sphere
\c{kh2}. The authors of \c{kh2} obtained that $\phi^4$ model of
this theory (nlgYM$_2$) on sphere has third order phase
transition. The authors of \c{kh3} have studied the large-$N$
limit of YM$_2$ and some features of it on cylinder.

The scheme of this paper is the following. In section 2 we briefly
review the large-$N$ limit of (nlgYM$_2$)theories for $U(N)$ gauge
group. In section 3, we study the phase structure of the theory
for $\phi^{2k}$ in all order and $\phi^2+\frac{2\alpha}{3}\phi^3$
models on arbitrary compact manifold with $g\neq 1$.

{\section{Preliminaries}}

The partition function of two dimensional  nonlocal generalized
Yang- Mills theories (nlgYM$_2$) on a compact manifold $\Sigma_g$
with genus $g$ and area $A$ is given by the exact formula
\c{{kh1},{kh2}} as: \be\l{1}
 Z_{\Sigma_g}(g,A)= \sum_R d^{2-2g}\exp\{\omega
[-A\Lambda(R)]\},
 \ee
where  $R$'s label the irreducible representation of the gauge
group, $d_R$ is the dimension of the representation $R$ and
$\Lambda (R)$ is
 \be \l{2}
  \Lambda (R) =
\sum_{k=1}^p\frac{\alpha_j}{N^{k-1}}C_k(R). \ee
 Here $C_k$ is the k'th Casimir of gauge group, $\alpha_j$'s are
 arbitrary constant. We consider the case that the gauge group is
 $U(N)$. The representation of this gauge group are labelled by
 $N$ integers $n_i$ satisfying $n_i\geq n_j (i\leq j)$ and it is
 found that
\bea \l{3}
 d_R&=& \prod_{1 \leq i \leq j \leq N} (1+\frac{n_i-n_j}{j-i}),\\
 C_k(R) &=& \sum_{i=1}^{N}[(n_i+N-i)^k - (N-i)^k].
 \eea
Now we redefine the function $\omega$ as: \be\l{5}
 -N^2V[A\sum_{k=1}^p \alpha_k {\hat C_k}(R)] := \omega [-A\Lambda(R)],
\ee where \be\l{6}
 {\hat C_k}(R) = \frac{1}{N^{k+1}}\sum_{i=1}^N (n_i+N-i)^k.
 \ee
 In the large $N$- limit, the above summation is replaced by a
 path integration over the continuous  function \c{do}
\begin{equation}\l{7}
\phi(x) =-n(x) + x - 1,
\end{equation}
 where
\be\l{8}
 0\leq x:=\frac{i}{N}\leq 1 \hspace{2cm} {\rm and} \hspace{2cm}
n(x):=\frac{n_i}{N}.
 \ee
The partition function can be rewritten as:
\begin{equation}\l{9}
Z_{\Sigma_g}[\phi (x)] = \int D\phi (x) \exp{\{-N^2S(\phi)\}},
\end{equation}
where
\begin{equation}\l{10}
S(\phi ) = V{\Biggr (}A \int _{0}^{1} W[\phi (x)] dx{\Biggl )} +
(1-g)\int_{0}^{1} dx \int_0^1 dy \log|\phi (x)- \phi (y)|,
\end{equation}
and \be\l{11}
   W(\phi ) := \sum_{k=1} (-1)^k \alpha_k \phi^k.
\ee Introducing the density function as $ u(\phi )
:=\frac{dx(\phi)}{d\phi }$ \c{do}, it is seen that it satisfies
\be\l{12}
 \int_{b}^{a} u(z) dz = 1,
  \ee
where $[b, a]$ is the interval corresponding to values of $\phi
(x)$ and also  the condition $n_i\geq n_j$ demands
 \be\l{13}
 u(z) \leq 1.
\ee
 As $N\rightarrow \infty$, only the configuration of $\phi$
 contributes to the partition function that minimizes $S$. To find
 this representation we put variation of $S$ with respect to
 $\phi$ equal to zero. So one can arrive at \c{kh2}
\begin{equation} \l{14}
h(z) = P\int_{b}^a \frac{u(z')dz'}{z - z'},
\end{equation}
where $P$ indicates  the principal value of integral and \be\l{15}
 h(z) = \frac{{\hat A}}{2(1-g)} W'(z),
  \ee
  and
\bea\l{16}
{\hat A} &:=& AV'{\biggr\{}A \int_0^1W[\phi (x)]dx {\biggl \}}, \nonumber \\
&=& AV'{\biggr\{}A \int_b^a W(z)u(z)dz{\biggl \}}.
 \eea
The free energy of the theory is defined as
\be\l{17}
 F := S|_{\phi_{cla.}},
 \ee
It is seen that
 \be \l{18}
 F'(A)= \frac{{\hat A}}{A} \int_{b}^au(z)W(z)dz.
 \ee
Using the standard method of solving the integral equation
(\r{14}), the density function $u(z)$ is obtained in terms of the
parameters $a$ and $b$. One can arrive at \c{kh2}. \be \l{19}
u(z)=\frac{\sqrt{(a-z)(z-b)}}{\pi}\sum_{n,m,q=0}^{\infty}\frac{(2n-1)!!(2q-1)!!}{2^{n+q}n!q!
(n+m+q+1)!}a^{n}b^{n}z^{m}h^{(n+m+q+1)}(0),
 \ee
and the values of $a$ and $b$ are determined from (\r{12}) and
\be\l{20} \sum_{n,q=0}^{\infty}\frac{(2n-1)!!(2q-1)!!}{2^{n+q}n!q!
(n+q)!}a^{n}b^{n}h^{(n+q)}(0)=0.
 \ee
The density function, $u(z)$ found from (\r{19}), depend on the
modified area $\hat{A}$, and therefore $A$. As $A$ increases, a
situation is encountered where $u$ exceeds 1. So, the density
function $u(z)$ violates the condition (13) for $A$'s larger than
some critical value $A_{c}$. $A_{c}$ is the value of $A$ at which
the maximum of $u$ becomes 1 ($u_{max}(A_{c})=1$). The region $A<
A_{c}$ is called the weak coupling phase (WCP) regime, and the
region $A>A_{c}$ is called the strong coupling phase (SCP) regime.

\section{Phase transition of nlgYM$_2$ on a compact manifold with
$g\neq 1$}

\subsection{$\phi^{2k} (k>1)$ model.}

In this case $W(\phi)$ is an even function of $\phi$, therefore
the density function in WCP regime, $u_w(z)$, is even, then
$b=-a$. So by rewriting (\r{19}), we have \be\l{21}
 u_w(z) =k\eta\sqrt{a^2-z^2}\sum_{n=0}^{\infty}\frac{(2n-1)!!}{2^nn!}a_k^{2n}z^{2k-2n-2},
\ee where $\eta = \frac{\hat{A}}{1-g}$ and $a_k$ is obtained from
(\r{12}) as: \be\l{22} a_k =
{\Biggr[}\frac{2^k(k-1)!!}{(2k-1)!\eta}{\Biggl ]}^{\frac{1}{2k}}.
\ee Equation (\r{21}) has three extremum points at $z=0$  and
$z_{1,2} = \pm a_k\sqrt{\zeta_k}$ \c{Ali}, in which $\zeta_k$ is
independent of $a_k$ and is determined from \be\l{23}
\sum_{n=0}^{k-2}\frac{(2n-1)!!}{2^{n+1}(n+1)}\zeta_k^{-(n+1)} = 1.
\ee By institute (\r{22}) in (\r{21}), we have
 \be\l{24}
u_{w}(z)=\eta^{\frac{1}{2k}}f(k)
 \ee
where $f(k)$ is independent of $\eta$($\hat{A}$or $A$). The value
of $A_{c}$ is obtained from \be\l{25} u_{w}(z_0)=1,
 \ee
 hence $z_0$'s are those extremum points, which $u_w(z)$ is
 maximum. One can arrive at
 \be \l{26}
 A_c V'(\frac{\hat{A_c}}{A_c})= \frac{1-g}{f(k)^{\frac{1}{2k}}}.
 \ee
 It is also easy to obtain $u''_w(z_0)$. Using $u'_w(z_0)=0$, one
 can see
  \bea\l{27}
u''_w(z_0) &=& \frac{k\eta z_0
a^{2k-3}(2k-1)}{\pi\sqrt{a^2-z_0^2}}{\Biggr [}\sum_{n=0}^{k-2}
\frac{(2n-1)!!z^{2k-2n-5}}{2^{n+1}(n+1)!a^{2k-2n-5}} -
(2k-1)a^{3-2k}z_0^{2k-3}{\Biggl ]}, \nonumber \\
&=&-\frac{k\eta
a^{2k-2}(2k-1)}{\pi\sqrt{a^2-z_0^2}}\sum_{n=0}^{k-2}
\frac{(2n-1)!!z_0^{2k-2n-4}}{2^nn!a^{2k-2n-4}}.
 \eea
 When $g>1$, then $\eta $ is negative, so that the $\phi^{2k}$
 model has no phase transition on compact manifold with $g>1$.
 But for the case $g=0$ ($\eta = \hat{A}$), (\r{27})  is clearly
 negative and density function  in WCP regime, $u_w(z)$, has a
 minimum at $z=0$ and two absolute maxima at $z_{1,2} = \pm a_k
 \sqrt{\zeta_k}$. Thus all $\phi^{2k}$'s models has phase
 transition  only on sphere. For areas slightly  more than the critical
 area, we can write the density function in WCP, $u_w(z)$, as:
 \be\l{28}
 u_w(z) = u_w(z_{1,2}) - {1\over 2}|u''_w(z_{1,2})|(z-z_{1,2})^2.
 \ee
 In the adjacent of critical point, $A_c$, $u_w(z) \geq 1$, and it
 is found the point $z'$ which satisfying $u_w(z') = 1$.
 Then
 \be\l{29}
 |z'-z_{1,2}| = \sqrt{\frac{2}{|u''_w(z_{1,2})|}}\xi,
 \ee
 where by using of (\r{24}) $\xi$ is
 \be\l{30}
 \xi = u_w(z_{1,2}) - 1 = \hat{A}^{\frac{1}{2k}}f(k) - 1.
 \ee
By the same procedure which used in \c{k1}, we can obtain that the
difference of free energy in SCP and WCP regime for nlgYM$_2$ is
\be\l{31}
 F_s-F_w \simeq \xi^3.
  \ee
One can expand $\xi(A)$ about $A_c$ as:
 \be\l{32}
  \xi(A) =
\xi(A_c) + \xi'(A_c) ( A-A_c) + \ldots , \ee where \be\l{33}
\xi'(A_c) = \frac{1}{2k}( \frac{d\hat{A}}{dA})_c\frac{1}{A_cV'(
\frac{\hat{A}_c}{A_c})}.
 \ee
So that, for the case $(\frac{d\hat{A}}{dA})_c \neq 0$, $\xi'(A)$
is nonzero and $\xi(A_c) = 0$, then
 \be\l{34}
 F_s - F_w = \beta
(A-A_c)^3.
 \ee
Hence $\beta$ is a constant which independent of modified area of
manifold, $\hat{A}$, ( or $A$). Therefore we conclude  that the
nlgYM$_2$ theories for all order of $W(\phi)= \phi^{2k}$ models
have third order phase transition only on sphere ($g = 0$).

\subsection{$W(\phi) = \phi^2 + \frac{2\alpha}{3}\phi^3$ models}

We  see that in the some nonlocal generalized case the constraint
(\r{13}) can be satisfy also for negative $\eta =
\frac{\hat{A}}{1-g}$ by an appropriate choice of the coupling
constant $\alpha_k$, which implies the existence of phase
transition at higher genera except the torus. Starting from
(\r{19}) for $W(\phi) = \phi^2 + \frac{2\alpha}{3}\phi^3$ model
and obtain
 \be\l{35} u_w(z)= \frac{\eta}{\pi}\{ 1+ {\alpha \over
2}(a+b) + \alpha z \} \sqrt{(z-b)(a-z)}, \ee and the value of $a$
and $b$ are determined from \bea\l{36}
 (a+b)(1+\frac{\alpha}{2}(a+b) ) + \frac{\alpha}{4}(a+b)^2 &=& 0, \\
\eta(a-b)^2(1 + \alpha(a+b)) &=& 8.
\eea
  By solving these equations,
we obtain \bea\l{38}
 b&=& \frac{2\lambda - 1}{2\alpha} - 2
\sqrt{\frac{\lambda}{\eta}},\\
a &=&\frac{2\lambda - 1}{2\alpha} + 2 \sqrt{\frac{\lambda}{\eta}},
\eea with
 \be\l{40}
  \lambda^3 - \frac{1}{4}\lambda +
\frac{\alpha^2}{2\eta} = 0 .
 \ee
 By substitute (\r{38}) and (39) in (\r{35}), we have
\be\l{41} u_w(y) = \frac{\eta}{\pi}\{ 2\lambda + y\}
\sqrt{\frac{4\lambda}{\eta} - y^2}, \ee where \be\l{41} y=z-
\frac{2\lambda - 1}{2\alpha}. \ee The density function $u_w(y)$,
(\r{41}), has two extremum points as:
\bea\l{43} y_1 &=& (Q - 1) \frac{\lambda}{2}, \nonumber \\
y_2 &=& -(Q +1)\frac{\lambda}{2}, \eea here $Q = \sqrt{1 +
\frac{8}{\lambda\eta}} > 1$. At the negative $\eta$ ($g > 1$), we
should take the $\lambda <0$ solution which exists for $\eta <
-6\sqrt{3}\alpha^2$. So from equations (\r {38}), (39) and (\r
{42}) we conclude that $y$ take the values in interval
$[-2\sqrt{\frac{\lambda}{\eta}} , 2\sqrt{\frac{\lambda}{\eta}}]$,
and the only extremum point which repose in
$[-2\sqrt{\frac{\lambda}{\eta}} , 2\sqrt{\frac{\lambda}{\eta}}]$
is $y_1$. It is clearly seen that $u_w(y)$ has an absolute maximum
in $[-2\sqrt{\frac{\lambda}{\eta}} ,
2\sqrt{\frac{\lambda}{\eta}}]$ at $y_1$. So that \be\l{44}
u''_w(y_1) < 0.\ee From (\r{41}), we have \be \l{45} u_w(y_1) =
\frac{\lambda\eta}{2\pi}(Q +3) \sqrt{\frac{4\lambda}{\eta} -
\frac{\lambda^2}{4}(Q - 1)^2}. \ee $\eta_c$ and therefore the
critical value for  modified  area of manifold, $\hat{A_c}$,( or
$A_c$) is determined by $u_w(y_1) = 1$, and it is found that
\be\l{46} \frac{\lambda_c\eta_c}{2\pi}(Q_c +3)
\sqrt{\frac{4\lambda_c}{\eta_c} - \frac{\lambda_c^2}{4}(Q_c -
1)^2} = 1. \ee By making use of (\r{32}), we have
\bea\l{47}
\xi'(A_c) &=& \frac{d}{dA}[u_w(y_1) - 1]|_{A=A_c},
\nonumber \\
&=&{\Biggr[}1 - \frac{1}{2(1-
\frac{\lambda_c\eta_c(Q_c-1)^2}{16})}{\Biggl
]}\frac{d}{dA}\ln\hat{A}|_{A=A_c}, \eea where if
$\frac{d\hat{A}}{dA}\neq 0$, then $\xi'(A_c) \neq 0$. So by the
same procedure in the previous subsection, one can obtain
\be\l{48} F_s - F_w  = \gamma \xi^3. \ee
 By substitute (\r{47}) in (\r{32}) and then in (\r{48}), we have
 \be\l{49}
 F_s - F_w = \gamma' ( A - A_c)^3, \ee
 where $\gamma'$ is a constant which is independent  of $\hat{A}$
 and therefore of $A$, so the order of  transition of this
 model is 3.

\section{conclusion}

We study $\phi^{2k}$ and $\phi^2 + \frac{2\alpha}{3}\phi^3$ models
for nlgYM$_2$ theories, and obtained that all order of
$\phi^{2k}$model has phase transition only on sphere and the order
of this transition is three. Also by considering the $\phi^2 +
\frac{2\alpha}{3}\phi^3$ model of  nlgYM$_2$, we found that, this
theory has third order phase transition on compact manifold with
$g > 1 $ and there is no phase transition on sphere. Note that the
whole reasoning is independent of the number of points at them
$u_w$ attains its  absolute  maximum. It is clear that similar
situation prevails for the cases which $u_w$ has many absolute
maximum. In this case, one can easily  realize the WCP regime as
the same technique in Preliminaries section, and then  obtain the
phase transition of theory in the multi - critical points. Also
remark that the critical value $\eta_c$ for any particular model
is fixed, therefore, increasing $\hat{A}$ the genus increases
proportionally in order to keep fixed the critical number of
handles per area, $\eta_c$. So if we fixed the modified area
$\hat{A}$ of the surface, then with the increase of the genus the
number of multi - critical points decreases and for  $ g > 1 +
\frac{\hat{A}}{|\eta_c|}$ there is no phase transition.

\end{document}